\documentclass[5p]{elsarticle}
\usepackage[utf8]{inputenc}
\usepackage{amsmath}
\usepackage{graphicx}
\usepackage[labelfont=bf]{caption}
\usepackage{wrapfig}
\usepackage{floatrow}
\usepackage{multicol}
\usepackage[exponent-product=\cdot]{siunitx}
\usepackage[version=4]{mhchem}
\usepackage{gensymb}
\usepackage{textcomp}
\usepackage{color}
\usepackage{indentfirst}
\usepackage{microtype}
\usepackage[colorinlistoftodos]{todonotes}
\usepackage[makeroom]{cancel}
\usepackage{booktabs}
\usepackage{ulem}
\usepackage{tabularx}
\newcolumntype{P}[1]{>{\centering\arraybackslash}p{#1}}
\begin{document}

\begin{frontmatter}

\title{Predictive Modeling of a Lithium Vapor Box Divertor in NSTX-U using SOLPS-ITER}

\author[First]{E.D. Emdee}
\author[First]{R.J. Goldston}
\author[Second]{J.D Lore}
\author[First]{X. Zhang}

\address[First]{Princeton Plasma Physics Laboratory, Princeton, NJ 08543, USA}
\address[Second]{Oak Ridge National Laboratory, Oak Ridge, TN 37831, USA}

\begin{abstract}

\indent The unmitigated heat flux in attached operation of a fusion power plant is predicted to be destructive to any solid divertor surface. Detachment, whereby the plasma pressure drops significantly before reaching the divertor target thus greatly reducing the heat flux and sputtering, will be necessary to ensure adequate lifetime of plasma facing components (PFCs). The lithium vapor box divertor aims to detach the divertor plasma via evaporating and condensing lithium surfaces. By evaporating lithium near or at the divertor plate and condensing it closer to the main chamber, a lithium vapor density gradient can be created. This density gradient ties energy losses to poloidal distance between the target and the detachment point. The radiation zone is then prevented from reaching the X-point as the lithium ionization rate decreases when the detachment front moves away from the divertor target. Here we present Scrape Off Layer Plasma Simulator (SOLPS) simulations of a lithium vapor box divertor using an NSTX-U equilibrium and PFC geometry. The parameters for the core boundary conditions, gas puff intensity, and heat and particle transport coefficients are chosen to match experimental values. Acceptable agreement with experimental Scrape-Off Layer (SOL) widths is found, indicating a reasonable choice of transport coefficients. In predictive simulations, lithium is added via evaporation at the target. Predictions for peak heat fluxes and upstream impurity concentration are given for a variety of evaporation rates. Target electron temperature is predicted to be able to be reduced to recombination levels (below 1 eV) for lithium evaporation rates of $1\cdot10^{23}$ Li/s, indicating detachment. Peak heat flux at the lower outer target could be reduced by as much as a factor of six while maintaining upstream lithium fractions below 2$\%$. The prevention of lithium from reaching the midplane is shown to be due to an increase in frictional forces acting on the lithium from a deuterium gas puff. Lithium is also shown to be redeposited close to the evaporator which is favorable for initial tests and future capillary porous systems. 
\end{abstract}

\begin{keyword}
divertor \sep lithium \sep vapor box \sep detachment \sep SOLPS-ITER \sep NSTX-U 

\end{keyword}

\end{frontmatter}
\section{Introduction}
The unmitigated SOL heat flux in a demonstration fusion reactor is predicted to be far beyond what any solid divertor could handle~\cite{SOLmodel,Kessel15}. Accordingly, more sophisticated divertors are integral to any successful design of a fusion reactor. Some requirements for a successful divertor are the reduction of peak target heat flux to below the engineering limit of the divertor surface and limiting impurity production to allow for strong core performance. Divertor detachment using gaseous seed impurities has generally succeeded in reducing the heat flux and sputtering at the target but has had more difficulty in preventing the seed impurities from reaching the main plasma~\cite{Potzel,AWleonard}. Typically, X-point MARFEs can form, causing significant radiation within the separatrix and reducing pedestal performance~\cite{MARFEs,MARFEdetach}. The goal of this research is to design a means to induce divertor detachment in a controlled manner by localizing radiation sources and impurities while achieving large heat flux reductions.

The lithium vapor box is a divertor design that seeks to control the detachment front by taking advantage of differential pumping~\cite{LVBpaper}. By evaporating lithium close to the divertor target and placing condensing surfaces between the target and the X-point, a vapor density gradient can be created~\cite{NMEvbpaper,EmdeePSI}. As the detachment front moves towards the X-point, less lithium will be at the ionization front, causing less energy loss, thus preventing further movement of the detachment front towards the X-point. Thus a natural feedback control is created which typical gas injections, such as neon or nitrogen, struggle to obtain. Lithium, being low-Z, will be minimally intrusive in case of divertor leakage; however, dilution is a primary metric of the design of a lithium vapor box divertor nonetheless.

Previous modeling on the lithium vapor box demonstrated its feasibility in reducing heat fluxes~\cite{UEDGE} as well as natural detachment front feedback control via the lithium vapor gradient~\cite{EmdeeNuclFus}. Additional modeling is now required to answer remaining questions about the vapor box, such as how to control upstream lithium concentration. The coupled fluid-Monte Carlo code SOLPS-ITER is used to model NSTX-U edge transport. The SOLPS profiles are matched with experimental data to calibrate the cross-field diffusion coefficients before lithium evaporation is introduced at the divertor plates for predictive modeling. Fueling puffs are shown to have a strong effect on upstream lithium fraction through a `puff and pump' mechanism whereby the deuterium friction force acting on the Li$^{+}$ is increased, reducing the amount of lithium escaping upstream~\cite{PuffAndPump}. This paper explores the control of the upstream impurity fraction, momentum balance, and the flow of energy and lithium in a device with a lithium vapor box.

\begin{figure}
\floatbox[{\capbeside\thisfloatsetup{capbesideposition={right,center},capbesidewidth=0.45\linewidth}}]{figure}[\FBwidth]
{\caption{A poloidal cross section of the NSTX-U equilibrium. Deuterium gas puffs already operating in the experimental shot are shown in dark green near the inner and outer midplane. A deuterium gas puff in the Private Flux Region (PFR) for controlling lithium fraction and a lithium evaporation region at the target are also added, shown in orange and red respectively (color online).}\label{fig:NSTXequilibrium}}
{\includegraphics[width=0.75\linewidth]{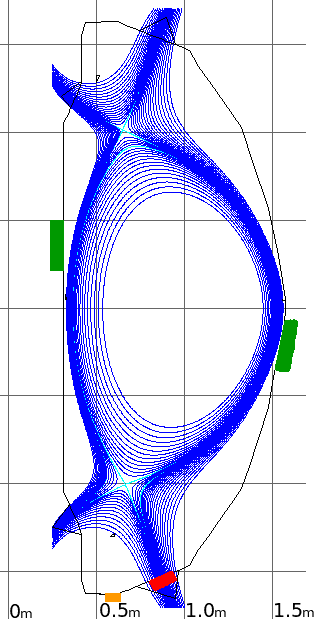}}
\end{figure}

\section{SOLPS-ITER}
SOLPS-ITER couples the fluid code B2.5 with the Monte Carlo neutral transport code EIRENE~\cite{SOLPS,bonninSOLPS}. B2.5 solves Braginskii-type fluid plasma transport equations to deduce ion fluid density, velocities and temperatures. The ion parallel momentum balance equation for a given lithium charge state is solved including thermal and friction forces with the main ions, other impurity ions, and the electrons. The form of the friction and thermal forces used in the 3.0.6 version of SOLPS-ITER is described by Sytova et al.~\cite{Sytova}. These forms of the impurity forces are more accurate than previous models for \(z_\mathrm{eff}>1.5\) plasmas, which is relevant for some of the simulations presented here. EIRENE is employed to model neutral lithium transport as well as atomic and molecular neutral deuterium transport against the plasma background provided by B2.5.

\subsection{Simulation Parameters}\label{parameters}

NSTX-U equilibrium 204202 at \SI{461}{ms} is used here for predictive vapor box modeling, with cross-field transport coefficients determined from experimental data. This shot had a disconnected double null with \SI{0.68}{MA} of plasma current, \SI{\sim2}{\mega\watt} of input power, and a dR$_{sep}\sim$ \SI{4}{mm}. A low field side and a high field side gas puff were on during the shot and included in the simulation. These puffs were set to a D$_2$ efflux of \SI{3.6e20}{D_2\per\second} and $4.8\cdot10^{20}$ D$_2$/s respectively. The positions of the gas puffs are noted in Figure \ref{fig:NSTXequilibrium}. The SOLPS modeling did not include carbon impurities, however they are estimated to have a small effect due to low concentrations as this was a low power L-mode discharge. A lithium gas puff, emulating an evaporator, and an additional deuterium gas puff in the private flux region were added for the predictive modeling, as shown in Figure \ref{fig:NSTXequilibrium}. The lithium evaporator had a length of \SI{11.8}{cm} in the poloidal plane roughly centered on the strike point, corresponding to a total area of \SI{0.64}{\square\meter}. The evaporated lithium efflux was scanned, corresponding to operating the lithium evaporator at different temperatures. The effluxes given and the corresponding temperature of the evaporator is shown in Table \ref{tab:temps}. The lithium evaporator temperature is calculated as described in previous work from experimental lithium equilibrium pressure \cite{EmdeePSI,handbook}.

\begin{table}
\begin{center}
\begin{tabular}{|P{0.35\linewidth}|P{0.35\linewidth}|}
\toprule
 \textbf{Lithium Efflux} & \textbf{Evaporator Temperature}\\
\midrule
$1 \cdot 10^{21}$ Li / s & 431\degree C \\
 \hline
$3 \cdot 10^{21}$ Li / s & 463\degree C \\
\hline
$1 \cdot 10^{22}$ Li / s & 501\degree C \\
\hline
$3 \cdot 10^{22}$ Li / s & 539\degree C \\
\hline
$6 \cdot 10^{22}$ Li / s & 563\degree C \\
\hline
$1 \cdot 10^{23}$ Li / s & 586\degree C \\
\bottomrule
\end{tabular}
\caption{A table of the lithium effluxes from the evaporator used and the corresponding temperature for the given evaporator area.}
\label{tab:temps}
\end{center}
\end{table}

\noindent As a point of reference, $1\cdot10^{23}$ Li/s represents 1.15 g/s of lithium, a significantly higher lithium injection rate than typical NSTX wall conditionings of approximately 1 mg/s using LITER devices prior to discharges \cite{gatesNSTX}. The amount of lithium deposited by the LITER devices ranged from a few milligrams to over \SI{2}{g} prior to a shot while the vapor box would run over the course of a five second shot time. Accordingly, the total amount of lithium injected by the vapor box during the shot would be on the same order as the highest pre-discharge lithium conditioning by the LITERs.

2 MW of input power was given as a boundary condition to the core interface, equally distributed between the electrons and ions. The B2.5 SOL outer edge was given a logarithmic derivative boundary condition consistent with a fit to the experimental data. Any lithium particle flux to this artificial boundary was assumed to be absorbed and any deuterium ion flux was assumed to be reflected back as a neutral deuterium. This particle absorption and reflection have little effect on the simulation as the particle flux to the outer radial boundary of the simulation was small. 

The SOLPS radial particle and thermal diffusion coefficients were adjusted to match outer mid-plane experimental data. The iteration scheme implemented is described by Canik et al. \cite{canik}. Experimental values are from Thomson scattering measurements. While the resulting agreement with the electron temperature and density profiles is not perfect, this method gives reasonable agreement with the expected heat flux channel, as shown in section \ref{matching}.

\subsection{Momentum Equation}

The parallel momentum equation used for species $a$ within the simulations shown is given by \cite{manual}.

\begin{align}\label{eq:mom_new}
&m_a \frac{\partial n_a V_{\parallel a}}{\partial t} +
\frac 1{h_z\sqrt g} \frac{\partial}{\partial x} \left( \frac{h_z\sqrt g}{h_x} \Gamma_{ax}^m \right)\notag\\ &+
\frac 1{h_z\sqrt g} \frac{\partial}{\partial y} \left( \frac{h_z\sqrt g}{h_y} \Gamma_{ay}^m \right) +
\frac{b_x}{h_x}\frac{\partial n_a T_i}{\partial x} \notag\\&+
Z_a e n_a \frac{b_x}{h_x} \frac{\partial \Phi}{\partial x} + \left( \nabla \cdot\overset\leftrightarrow\pi_a^{\parallel} \right)_{\parallel}
= \notag \\&S_{CF_a}^m + S_{fr_a}^m + S_{\mathrm{Therm}_a}^{m} + S_{I_a}^m \notag\\&+ S_{R_a}^m + S_{CX_a}^m + S_{AN_a}^m + S_{\mathrm{EIRENE}_a}^m
\end{align}
where $\Gamma_{ax}^m$ is the momentum flux in the poloidal direction and  $\Gamma_{ay}^m$ is the momentum flux in the radial direction. $h_{x,y} = \frac{1}{|\nabla x,y|}$ and $h_z=2 \pi R$. $\sqrt{g}$ is the volume element. $\Phi$ is the electric potential. $\overset\leftrightarrow\pi$ is the viscous stress tensor.

\begin{figure}
\centering
\includegraphics[width=1.0\linewidth]{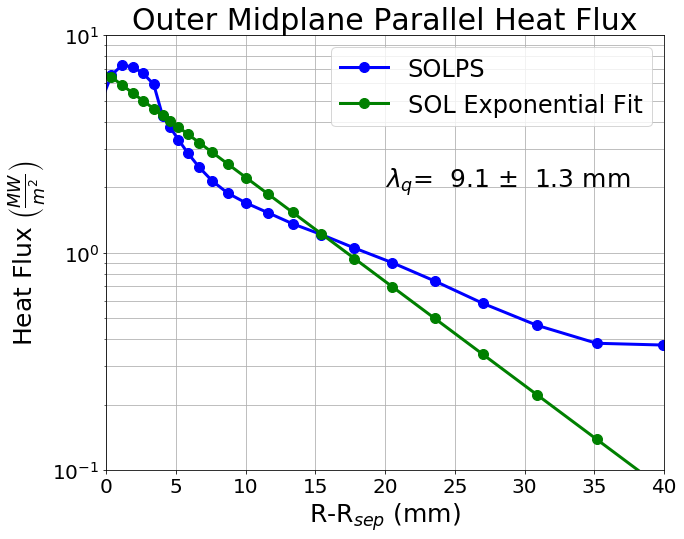}
\caption{NSTX outer midplane parallel heat flux from SOLPS and exponential fits to the full SOL and to just the outer SOL.}
\label{fig:SOLwidthFits}
\end{figure}

The $S^m$'s are the various momentum sources: the coriolis force, friction forces, thermal forces, ionization, recombination, charge exchange, anomalous current, and plasma-neutral interactions from EIRENE, respectively. The other variables have their usual meaning. Analysis of the sources determines that the friction force due to the deuterium on the lithium is generally the primary source of momentum towards the divertor plate for the lithium while the thermal force is the dominant source of momentum towards the main chamber.

\section{Results}\label{results}
\subsection{Profile Matching}\label{matching}

A fit to the outer midplane (OMP) parallel heat flux gives a SOL power width consistent with past experiments on NSTX, which found SOL power widths in the range of \SIrange{3}{14}{\milli\meter} for lower single-null H-mode shots \cite{NSTXwidth}. Experimental heat flux width measurements are not available for this NSTX-U shot, necessitating comparison with scaling relations. The Eich SOL power width scaling \cite{Eich}, when using data from MAST and NSTX single-null H-modes (regression $\#$15), yielded $\lambda_q\sim$\SI{3.0}{mm} using the parameters of this shot. The Goldston heuristic drift model would predict a SOL width at the OMP of \SI{2.3}{mm} for low gas puff single-null H-modes~\cite{HDmodel,EichASDEX}. L-mode Heat flux width scaling from single-null MAST shots \cite{LmodeScaling} using the parameters of this NSTX-U shot indicates a $\lambda_q^{mp}\approx\SI{13.8}{mm}$.

\begin{figure}
\centering
\includegraphics[width=1.0\linewidth]{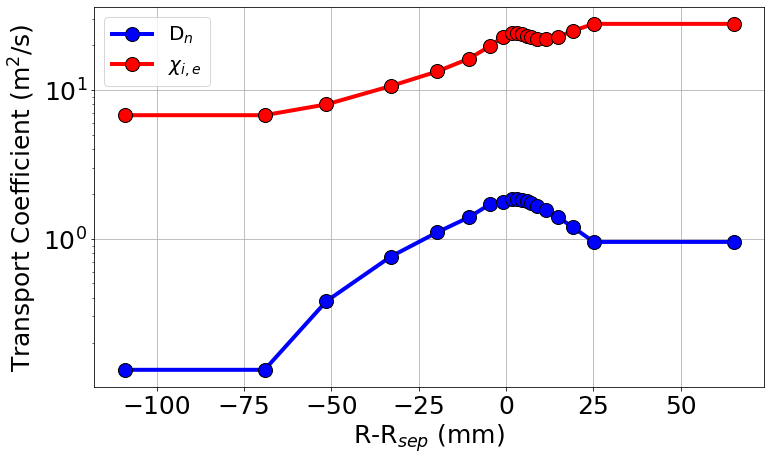}
\caption{Derived transport coefficients from the fitting algorithm described in section \ref{parameters}.}
\label{fig:TCs}
\end{figure}

Importantly, one should keep in mind that this shot is a double null configuration with a dR$_{sep}\sim$\SI{4}{mm}, making it difficult to compare with scaling relations fit to single null discharges. Looking at Figure~\ref{fig:SOLwidthFits}, one can see that the first few radial zones beyond R-R$_{sep}=0$ clearly do not conform to exponential behavior due to the inter-separatrix region (R-R$_{sep}\sim$0-4mm). Fitting a single exponential to the full SOL yielded a $\lambda_q$ of 9.1$\pm$\SI{1.3}{mm}. Thus the SOL width produced by SOLPS could be said to be similar to typical spherical tokamak SOL widths for L-mode discharges, though it does appear the transport coefficients produce a narrower SOL than a typical L-mode would provide. The transport coefficients derived in the fitting process are shown in Figure \ref{fig:TCs} and taken to be reasonable for an \mbox{NSTX-U} plasma. These transport coefficients are then used to give predictions for a lithium vapor box divertor in \mbox{NSTX-U}.

\begin{figure}
\centering
\includegraphics[width=1.0\linewidth]{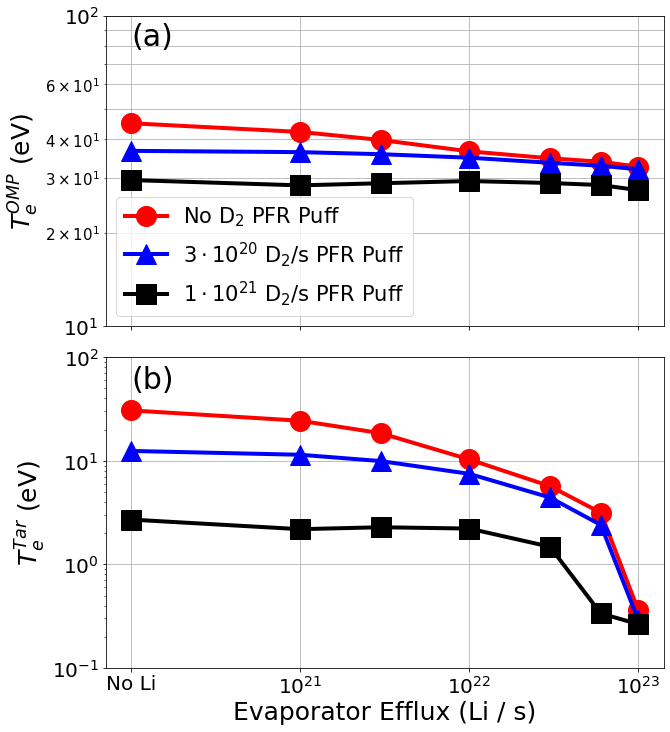}
\caption{SOLPS separatrix electron temperatures at the (a) outer midplane and (b) lower outer target for a variety of PFR D$_2$ puff intensities and Li evaporator effluxes}
\label{fig:electronTemps}
\end{figure}

\subsection{Detachment Via Lithium Evaporation}\label{mainResults}

The rate of lithium evaporation as well as the intensity of the PFR D$_2$ puff were scanned in the simulations. The resultant electron temperatures at the separatrix are shown in Figure \ref{fig:electronTemps} for both the target and the OMP. It can be seen that for $1\cdot10^{23}$ Li/s (at 586 $\degree$C) the target temperature drops to \SI{\sim0.3}{eV} regardless of \ce{D2} puff intensity. \ce{D+} ion flux to the lower outer target is heavily reduced at this point, indicating that the plasma is fully detached regardless of the PFR D$_2$ puff. Signs of detachment are also present in Figure \ref{fig:ionization} where the ionization front is shown to lift off the target as lithium evaporation is increased.

It should be noted that attempting to detach the plasma solely with deuterium gas injection led to a collapse in the upstream plasma temperature as the plasma detached (not shown here). In fact, no solutions were found that detached solely with deuterium and had a T$_e^{OMP} >$ \SI{17}{eV}. Thus the lithium provides novel benefit to the scenario presented.

\subsection{Lithium Fraction Control}

\begin{figure}
\centering
\includegraphics[width=1.0\linewidth]{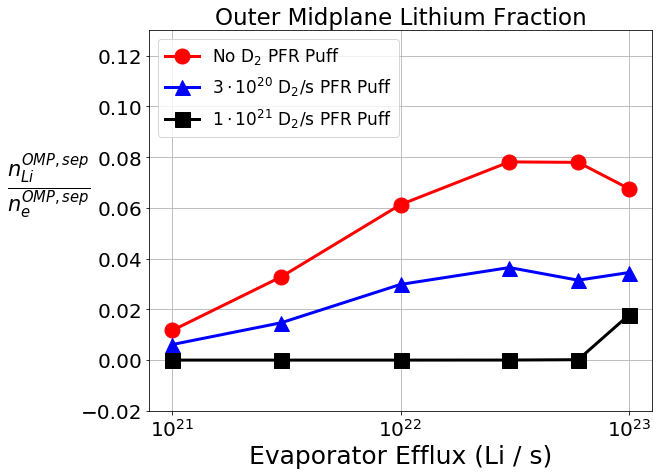}
\caption{Predicted lithium fractions at the outer midplane for different evaporator effluxes and \ce{D2} puff intensities. The lithium fraction is strongly dependent on the amount of \ce{D2} puffed in, allowing the lithium fraction to be effectively controlled.}
\label{fig:liFracs}
\end{figure}

The lithium fraction at the outer midplane is strongly dependent on the intensity of the D$_2$ puff. This is shown in Figure \ref{fig:liFracs}. One can see that without the PFR D$_2$ puff, the lithium ions can equal up to $\sim$8$\%$ of the electron density at the separatrix OMP. A subtlety of Figure \ref{fig:liFracs} is that the \(n_e\) at the separatrix OMP is also increasing since the transport coefficients are kept constant. For example, the case with 1$\times10^{23}$ Li/s evaporation and 1$\times10^{21}$ \ce{D2}/s gas injection has a 25$\%$ higher OMP separatrix electron density than the case with the same lithium evaporation but no deuterium injection. However, as one can see via Figure~\ref{fig:liFracs} and Figure~\ref{fig:impurityForces}, the total lithium content at the OMP decreases dramatically further than 25$\%$ with higher deuterium puffs. Thus, while the divertor target temperature can be strongly reduced without the \ce{D2} injection, the additional \ce{D2} injection allows for an upstream plasma with minimal fuel dilution. At \SI{6e22}{Li\per\second} and \SI{1e21}{D_2\per\second}, the target temperature is still low at \SI{\sim0.3}{eV} and the lithium fraction is virtually non-existent at the OMP, reduced from 8\% without the PFR \ce{D2} puff. For \SI{1e23}{Li\per\second}, the lithium fraction was able to be lowered to below 2\%. The physics behind the lithium fraction reduction is explored in section \ref{momentum}.

\subsection{Sources of Momentum}\label{momentum}


\begin{figure}
\centering
\includegraphics[width=1.0\linewidth]{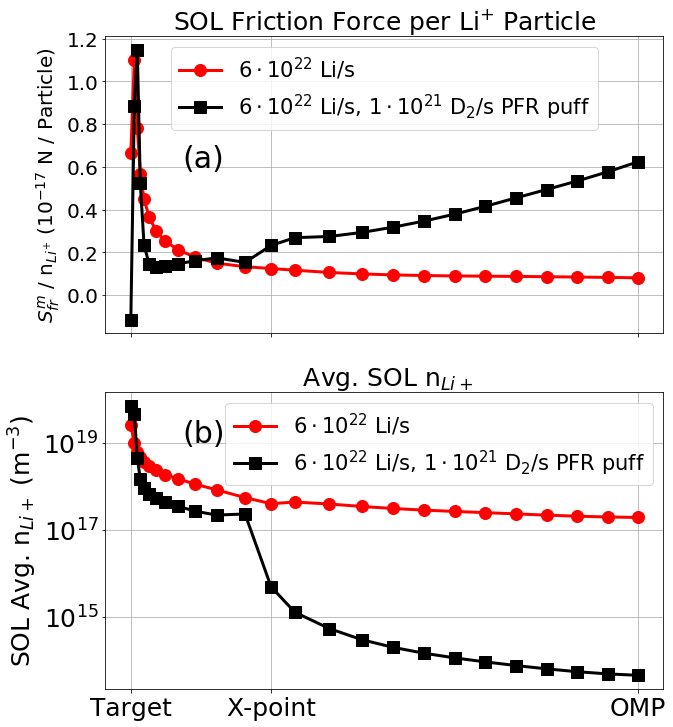}
\caption{(a) The average friction force acting on an Li$^{+}$ particle (b) The resulting density of Li$^{+}$ averaged across the SOL flux tubes. The sharp increase in the friction force (in the case with PFR \ce{D2} puffing) around the X-point is seen to coincide precisely with a large decrease in Li+ density.}
\label{fig:impurityForces}
\end{figure}

\begin{figure}[!t]
\centering
\includegraphics[width=0.97\linewidth]{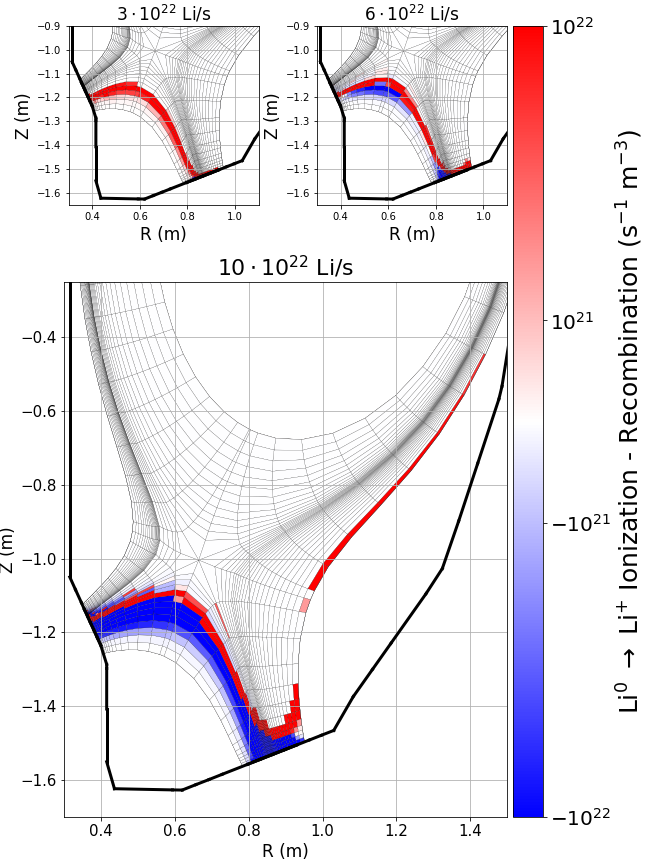}
\caption{Poloidal cross-section of the volumetric ionization rate for three cases with $1\cdot10^{21}$ \ce{D2}/s. At $1\cdot10^{23}$ Li/s the ionization front lifts fully off the target, indicating full detachment at this evaporation rate.}
\label{fig:ionization}
\end{figure}

To understand the variation of the separatrix OMP lithium fraction, $S^{m,Li+}_{fr}$, the friction force on the Li$^+$ fluid, was averaged across the SOL flux tubes and normalized by the amount of lithium in the SOL at a given poloidal index. The result is then plotted against the corresponding distance along the separatrix in Figure \ref{fig:impurityForces}. A clear distinction between the $1\cdot10^{21}$ \ce{D2}/s case and the case without PFR \ce{D2} puffing is seen in $S^{m,Li+}_{fr}$ around the X-point and further upstream. This explains the steep gradient in lithium ion density at the X-point. A steep lithium density gradient is also present near the target in the case with the deuterium puff in the PFR. This is due to a weaker thermal force from energy losses directly at the target, causing the friction force to more effectively screen lithium ions close to the target.

This is evidence of a `puff and pump' effect whereby the ion friction force pushes the impurities towards the divertor plate, ensuring the upstream plasma is uncontaminated. The highest lithium evaporation rate had non-negligible lithium at the OMP because the neutral lithium was able to escape into the main chamber, causing upstream ionization as shown in Figure \ref{fig:ionization}. Keeping in mind that the NSTX-U geometry used here represents a completely open divertor, the upstream ionization may not be a problem in a true vapor box divertor geometry. Exploring the effect of divertor closure and evaporator location will be the subject of future work.

\subsection{Condensed Lithium Distribution}\label{particleBalance}

\begin{figure}
\centering
\includegraphics[width=1.0\linewidth]{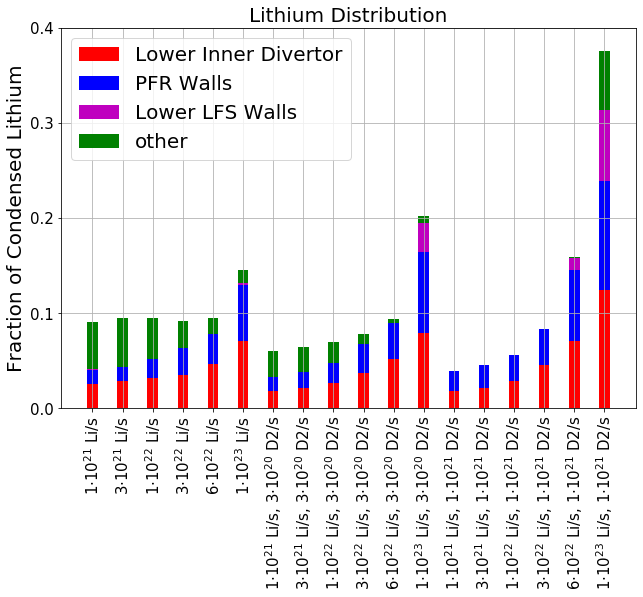}
\caption{The distribution of where the lithium ends up for each of the lithium cases, other than the lower outer divertor target. Across all cases most of the lithium ends up back at the lower outer divertor. The listed D$_2$/s is the PFR puff intensity. The ``other'' category primarily consists of flux to the upper outer divertor but also contain some lithium ending up at the higher LFS walls.}
\label{fig:liPumped}
\end{figure}

The lithium was condensed at any wall surface it came into contact with, as it would be for the low temperature NSTX-U walls. The distribution of where the lithium is pumped is of note, as this could inform the needs of a potential capillary porous system (CPS) for recirculation of lithium as discussed in previous work~\cite{EmdeeNuclFus,filtration,OnoLoop}. If the lithium is too spread out, the demands on a recirculation system may become large. Even without a recirculation system, the location of the settled lithium could inform chamber preparation for any first lithium vapor box test. Accordingly, the distribution of lithium not condensed at the lower outer target (the walls beyond the last poloidal cell index of the grid shown in Figure \ref{fig:ionization}) is given in Figure~\ref{fig:liPumped}.

\begin{figure}
\centering
\includegraphics[width=1.0\linewidth]{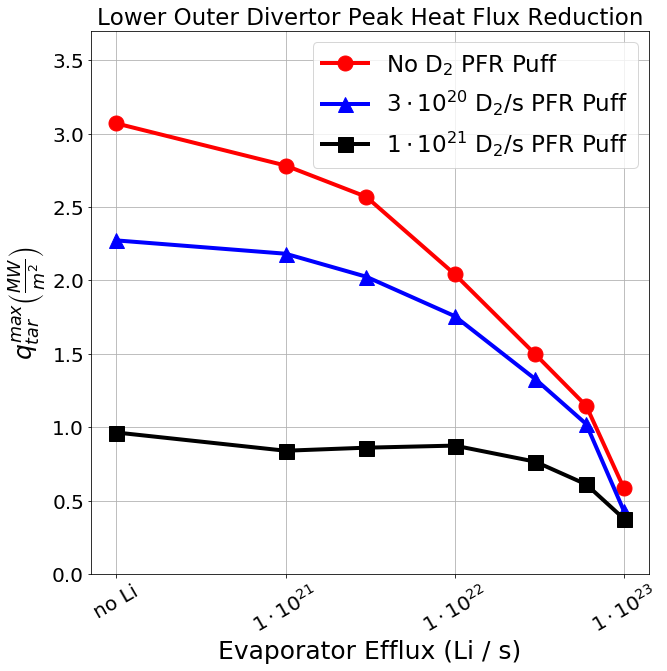}
\caption{The peak heat flux at the lower outer divertor for the different Li and D$_2$ effluxes. Peak heat flux was able to be reduced by $\sim80\%$ from the base case.}
\label{fig:qmax}
\end{figure}

It can be seen that most of the lithium ends up back at the lower outer divertor target (since no fraction in Figure \ref{fig:liPumped} ever exceeds 0.4). It should be noted that 85-95$\%$ of the lithium that hit the lower outer target hit the region with the evaporator. This exceeds the fraction of the target area the evaporator constitutes, which was $\sim62\%$. The high concentration of lithium flux towards the strike point is beneficial since the demand for recirculation will be low. Similarly, the need to remove lithium from the walls in any experiment without a built-in recirculation system would be less frequent. The lithium secondarily ended up at the PFR walls and the lower inner divertor. The `other' category in Figure \ref{fig:liPumped} typically consisted of the upper outer divertor though in some cases, especially those with high quantities of lithium, some lithium was condensed at the low-field side walls closer to the outer midplane.

\subsection{Peak Heat Flux Reductions}

Peak heat flux was able to be reduced by as much as a factor of 8 from the experimentally matched base case as shown in Figure \ref{fig:qmax}. This closely follows the reduction in target electron temperature seen in Figure \ref{fig:electronTemps}. This is opposed to attempting to detach entirely with deuterium, which was unable to decrease the heat flux much further than a factor of 3 without reducing upstream temperatures to be less than \SI{20}{eV}. With increased neutral injection (both deuterium and lithium) radiation and heat transfer to the neutrals is increased, thus moving the heat flux away from the targets. As an example, at $1\times10^{21}$ \ce{D2}/s, the heat transfer to the neutrals went from \SI{0.39}{MW} at no lithium evaporation to \SI{0.65}{MW} at $1\times10^{23}$ Li/s. The radiation for the same cases went from \SI{0.74}{MW} to \SI{1.02}{MW}. Under these low power conditions, the heat flux reduction due to line radiation and recombination/ionization energy loss is smaller than the heat flux reduction due to neutral excitation radiation with neutral excitation radiation making up $\sim$95$\%$ of the total radiation in the case with $1\times10^{23}$ Li/s and $1\times10^{21}$ D$_2$/s from the PFR. Of this neutral excitation radiation, 40$\%$ was due to neutral lithium, 45$\%$ was due to atomic deuterium, and the rest was due to molecular deuterium. This is consistent with other work on the lithium vapor box in a low power-density environment \cite{schwartz}. 

\section{Conclusions and Future Work}

The lithium vapor box has been shown to be effective at reducing the divertor target heat flux, achieving full detachment while maintaining low upstream lithium concentrations. The target temperature drops dramatically for modest drops in outer midplane temperature, indicating isolation of the main plasma from divertor cooling. Higher lithium evaporation rates saw ionization further upstream, causing non-negligible OMP lithium fraction. However, the lithium fraction at the OMP can be controlled via a fueling puff in the PFR. This is due to an increase in the deuterium friction force on the \ce{Li+}.

The lithium primarily condenses at the lower inner divertor target, the PFR and the lower outer divertor target, suggesting beneficial properties for initial lithium vapor box tests and ultimately lithium recirculation. The peak heat flux was able to be reduced by a factor of 8 from the experimentally matched base case, mostly because of neutral excitation and radiation. The effect of divertor closure, high power operation, and drifts along with the lithium vapor box's behavior in both NSTX-U and a pilot plant design with high temperature walls that do not accumulate lithium will be the subject of future research.

\section*{Acknowledgements}
\noindent This work is sponsored by DOE Contracts No. DE-AC02-09CH11466, DE-AC52-07NA27344, and DE-AC05-00OR22725. A.O.\ Nelson is thanked for many helpful conversations. J.A.\ Schwartz is thanked for his help on manuscript revision. The NSTX-U Team is thanked for access to experimental data. Data for this paper will be made available upon request.
\nocite{*} 

\end{document}